\documentclass[10pt,letterpaper]{article}

\usepackage[top=0.85in,left=2.75in,footskip=0.75in]{geometry}

\usepackage{amsmath,amssymb}

\usepackage{ulem}

\usepackage{dirtytalk}

\usepackage{changepage}

\usepackage[utf8x]{inputenc}

\usepackage{textcomp,marvosym}

\usepackage{cite}

\usepackage{nameref,hyperref}

\usepackage[right]{lineno}

\usepackage{microtype}
\DisableLigatures[f]{encoding = *, family = * }

\usepackage[table]{xcolor}

\usepackage{array}

\newcolumntype{+}{!{\vrule width 2pt}}

\newlength\savedwidth

\raggedright
\usepackage[aboveskip=1pt,labelfont=bf,labelsep=period,justification=raggedright,singlelinecheck=off]{caption}

\makeatletter
\renewcommand{\@biblabel}[1]{\quad#1.}
\makeatother

\usepackage{lastpage,fancyhdr,graphicx}
\usepackage{epstopdf}

\fancyheadoffset[L]{2.25in}
\fancyfootoffset[L]{2.25in}
\begin{document}
\vspace*{0.2in}

\begin{flushleft}
{\Large
\textbf{Hurricanes Increase Climate Change Conversations on Twitter
}
\newline
\\ Maddalena Torricelli\textsuperscript{1},
Max Falkenberg\textsuperscript{1},
Alessandro Galeazzi\textsuperscript{2},
Fabiana Zollo\textsuperscript{2,3},
Walter Quattrociocchi\textsuperscript{4},
Andrea Baronchelli\textsuperscript{1,5,*} \\
\bigskip
\small
\textbf{1} City University of London, Department of Mathematics, London EC1V 0HB (UK)
\\
\textbf{2} Ca’ Foscari University of Venice — Department of Environmental Sciences, Informatics and Statistics, Via Torino 155, 30172 Venezia (IT)
\\
\textbf{3} The New Institute Centre for Environmental Humanities, Dorsoduro 3911, Venezia (IT)
\\
\textbf{4} Sapienza University of Rome — Department of Computer Science, Viale Regina Elena, 295 — 00161 Roma (IT)
\\
\textbf{5} The Alan Turing Institute, British Library, London NW1 2DB (UK)
\\
\textbf{*} Corresponding author: abaronchelli@turing.ac.uk
\normalsize
\bigskip
}
\end{flushleft}

\section*{Abstract}
The public understanding of climate change plays a critical role in translating climate science into climate action. In the public discourse, climate impacts are often discussed in the context of extreme weather events. Here, we analyse 65 million Twitter posts and 240 thousand news media articles related to 18 major hurricanes from 2010 to 2022 to clarify how hurricanes impact the public discussion around climate change. First, we analyse news content and show that climate change is the most prominent non-hurricane specific topic discussed by the news media in relation to hurricanes. Second, we perform a comparative analysis between reliable and questionable news media outlets, finding that the language around climate change varies between news media providers. Finally, using geolocated data, we show that accounts in regions affected by hurricanes discuss climate change at a significantly higher rate than accounts in unaffected areas, with references to climate change increasing by, on average, 80\% after impact, and up to 200\% for the largest hurricanes. Our findings demonstrate how hurricanes have a key impact on the public awareness of climate change. 


\section*{Introduction}
\label{sec:intro}

Discussions around climate change are pervasive across environmental policy \cite{leiserowitz2007international, portner2022climate, fekete2021review}, political debate \cite{david2022public}, and public opinion \cite{shwom2015public}. 
Nonetheless, given the significant polarization around climate-change beliefs \cite{falkenberg2022growing}, it is imperative to fully investigate the factors which shape individual perspectives on this crucial issue. This challenge becomes all the more important in scenarios where attributing causality to climate change is complex, for instance in the case of hurricanes, where the role of climate change remains a contested matter \cite{pielke2005hurricanes,weinkle2018normalized}.
In this paper, we take a step towards addressing this need by asking how the impact of a hurricane on a local area affects public attention towards climate change.

Our analysis combines social media data from Twitter with a dataset of news media articles concerning hurricanes. Social media is known to play a pivotal role in facilitating discussions around climate change \cite{kirilenko2014public, pearce2019social,falkenberg2022growing,omodei2015characterizing,spaiser2022dare}.  In particular, social media may be contributing to growing polarization, especially in relation to views on climate change \cite{falkenberg2022growing}, with users confined to climate-sceptic or pro-climate action echo-chambers \cite{cinelli2021echo}. These echo-chambers attract politicians and users with opposed views, and often reference different news media outlets. 

The traditional news media plays a critical role in the public understanding of, and attention towards, climate change \cite{hart2015public, jamison2010climate}. Research has shown how news media coverage of climate change has evolved over the years \cite{pearce2019social}, particularly when discussing climate science or policy, with the personalization and dramatisation of climate news blamed for a lack of accurate and informative media coverage \cite{boykoff2007climate}. This is particularly problematic when discussing individual events where attribution to climate change is difficult, or where evidence for the role of climate change in driving, for instance, hurricanes is disputed \cite{pielke2005hurricanes,weinkle2018normalized}. 

These uncertainties are important since there is evidence that people's attitude towards climate change is influenced by extreme weather. For example, one study noted how even small variations in local weather temperature patterns can impact an individual's perception and discussion of climate change \cite{mumenthaler2021impact}.
When we discuss hurricanes, it is known that people's opinions on climate change can change if they are directly affected by one \cite{sloggy2021changing}. However, how their views change depends on the social and political context of the individuals and communities involved \cite{roxburgh2019characterising}. Some studies have shown that politicians frequently do not accept the role that climate change plays in extreme weather, or that little can be done to prevent such events. However, the public often blames governments for not doing enough to prevent or handle these situations \cite{lahsen2022politics}.

Here, we extend this literature by performing a comprehensive, longitudinal analysis on a large number of hurricanes to quantify the change in the online discussion around climate change when a region is impacted by a hurricane. To tackle this challenge, we rely on Twitter data as a proxy for public attention \cite{pearce2019social}. While Twitter users are not wholly representative of the general population, the platform is a useful tool for identifying trends and changes in public opinion, and has a disproportionate influence on the views of politicians and journalists \cite{jungherr2016twitter, mitchell2021news, jacobs2016social, klinger2015emergence, chadwick2017hybrid}. 

In the remainder of this paper, we first outline our three datasets: two Twitter datasets, totalling approximately 65 million tweets on climate change and on 18 of the most severe North Atlantic hurricanes from 2010 to 2022, and a dataset of news summaries which are referred to in the tweets about hurricanes. We provide a description of each dataset, and use a topic modelling approach to understand the themes discussed by the news media around hurricanes and climate change. In particular, we consider how content differs in its language depending on whether the news source is reliable or questionable. Finally, we reveal the local impact of hurricanes on the discussion of climate change using geo-located tweets, but reveal that there is a rapid decay in public attention in subsequent weeks. We end the paper by discussing our results and their implications for climate communication policies. 

\section*{Materials and methods}

\subsection*{Data}
\label{sec:data_collection}

In this Section, we introduce three distinct datasets related to hurricanes and climate change. The first dataset consists of tweets related to hurricanes, which were collected using the official Twitter API, referred to as the \textit{hurricane tweets} throughout the paper. This dataset consists of more than 36 million original tweets (i.e. excluding retweets) posted by more than 6 million users mentioning the 18 hurricanes between January 1, 2010 and December 31, 2021 whose names were retired, see Table~\ref{tab:metadata_hurricanes}. We only study hurricanes whose names have been retired to ensure that keyword searches for individual hurricane names refer to a single unique hurricane. We do not analyse content prior to 2010 due to a lack of meaningful tweet volume. 

Searching the hurricane tweets for URLs referencing known news domains, we collect more than $240$ thousand news articles that are referenced in the context of the 18 selected hurricanes. This set of news articles is used to perform the topic modelling described in Section \nameref{sec:top_mod}. 

Finally, our third dataset contains all original tweets using the term \say{climate change} on Twitter from January 1, 2010 to December 31 2021, totalling more than $29$ million tweets posted by more than 4 million users. In this dataset, approximately $2\%$ of tweets are geolocated at the state level. We refer to this dataset as the \textit{climate change tweets}.

\subsection*{Topic modelling}
\label{sec:top_mod}

To analyze the 240 thousand news articles referred to in the hurricane tweets we use a topic modelling approach. The topic modelling tool BERTopic \cite{grootendorst2022bertopic} extracts latent topics from a group of documents. It is well suited for analysing Twitter data where tweets are documents from whose texts the model can derive coherent themes, due to its ability to generate a vector representation of sentences while preserving their semantic structure \cite{egger2022topic, asgari2021topic}. The algorithm creates document embeddings using pre-trained transformer-based language models. It then produces topic representations by clustering embeddings using a class-based TF-IDF procedure \cite{claude2010tfidf}. This tool has proved to be effective in classifying topics from Twitter posts \cite{falkenberg2022growing,mekacher2023systemic}, including in relation to climate change \cite{falkenberg2022growing}.

For each news article, we use NewsGuard, a media reliability assessor, to classify a news source as reliable or unreliable. NewsGuard editors analyze news outlets based on nine journalistic criteria \cite{newsguard}. These criteria are used to assign a reliability score to each news outlet between 0 and 100. Outlets with a score lower than 60 are considered unreliable. Reliability scores from NewsGuard are known to be broadly in line with scores from other media reliability providers \cite{lin2022high}.

Finally, to compare the language used by reliable and unreliable news sources when discussing hurricanes we used the Shiftiterator package \cite{gallagher2021generalized}. This package creates word shift graphs that highlight which words contribute to understanding the differences between two texts. The comparison method is based on word frequency counts, and the proportion shift of each word is calculated by evaluating the probabilities that the word appears in each text. 

\subsection*{Geolocated tweet analysis}
\label{sec:stat_an}

To assess how discussions around climate change are affected by hurricane impacts, we count the number of tweets inside and outside the impact area of each hurricane in the period of a one month before the impact date and three months after it. We define a tweet \textit{in location} if its geolocalisation collocates it in the state affected by a specific hurricane; we define it \textit{out of location} otherwise. The areas of impact of each hurricane have been determined using \cite{smith2020us, world2022state}. As a result, we obtain, for each day, two distributions of the number of tweets for each hurricane for both the \textit{in location} and the \textit{out of location} regions.
The two \textit{in location} and \textit{out of location} counts of the number of tweets for each hurricane are normalised using the relative average number of tweets in the 30 days before the hurricane. Table \ref{tab:metadata_hurricanes} reports the counts for \textit{in location} and \textit{out of location} for all the hurricanes aggregated over one month before and one month after the impact respectively.

\begin{table}[ht!]
\centering
\begin{tabular}{|l|l|l|l|l|l|l|}
\hline
\textbf{Hurricane} &
\multicolumn{1}{|p{1.3cm}|}{\centering \textbf{In\\location\\(before)}} &
\multicolumn{1}{|p{1.3cm}|}{\centering \textbf{In\\location\\(after)}} &
\multicolumn{1}{|p{1.3cm}|}{\centering \textbf{Out of\\location\\(before)}} &
\multicolumn{1}{|p{1.3cm}|}{\centering \textbf{Out of\\location\\(after)}} &
\multicolumn{1}{|p{1.3cm}|}{\centering \textbf{Damage\\(billion USD)}} &
\textbf{Month}\\ \hline
Tomas & 0 & 0 & 65174 & 71163 &0.3 & Oct-Nov, '10\\
\hline
Irene & 55 & 88 & 53210 & 77436 &142.0 & Aug, '11\\
\hline
Sandy & 109 & 507 &  89519 & 213982 & 687.0 & Oct,'12\\
\hline
Ingrid & 18 & 30 & 122234 & 177436 &15.0 & Sept,'13\\
\hline
Erika & 184 & 205 & 281762 & 340991 &0.5 & Aug,'15\\
\hline
Joaquin & 54 & 78 & 346666 & 335414 &0.2 & Sept-Oct,'15\\
\hline
Matthew & 334 & 474 & 235957 & 265128 & 151.0 & Sept-Oct,'16\\
\hline
Otto & 13 & 6 & 301822 & 318113&0.2 & Nov,'16\\
\hline
Harvey & 483 & 1107 & 221326 & 344326 &125.0 & Aug-Sept,'17\\
\hline
Irma & 531 & 1086 & 237229 & 350830 & 772.0 & Aug-Sept,'17\\
\hline
Maria & 170 & 71 & 333506 & 212094 & 916.0 & Sept,'17\\
\hline
Nate & 25 & 19 & 307093 & 205139 &0.8 & Oct,'17\\
\hline
Florence & 54 & 65 & 187104 & 192252 &24.0 & Aug-Sept,'18\\
\hline
Michael & 310 & 624 & 180396 & 348129 & 251.0 & Oct,'18\\
\hline
Dorian & 697 & 986 & 341384 & 473079 &51.0 & Aug-Sept,'19\\
\hline
Laura & 144 & 442 & 149860 & 282564 & 191.0 & Aug,'20\\
\hline
Eta & 413 & 275 & 235765 & 219886 &83.0 & Oct-Nov,'20\\
\hline
Iota & 16 & 16 & 229882 & 211220 &14.0 & Nov,'20\\
\hline
\end{tabular}
\caption{\textbf{Number of tweets \textit{in location} and \textit{out of location} aggregated over one month before and after the impact of each hurricane.} The Damage column lists the cost of damages caused by each hurricane in US Dollars \cite{smith2020us}.}
\label{tab:metadata_hurricanes}
\end{table}

We analyse this data to assess how attention towards climate change varies before and after a hurricane impacts, in and outside the regions impacted. To compute statistics, we normalise counts for each hurricane and each day as
\begin{eqnarray}
    \hat{x}^{loc}=\frac{x^{loc}_{after} - c^{loc}_{norm}}{c^{loc}_{norm}}
\end{eqnarray}
where we indicate as $\hat{x}$ the fractional change in the tweet count after the impact ($x_{after}$) with respect to $c_{norm}$, the count normalisation on one month before the impact defined as
\begin{eqnarray}
    c^{loc}_{norm}=\frac{\sum_{day=1}^{30} x^{loc}_{before}(day)}{30};
\end{eqnarray}
where $x_{before}$ represents the tweet count before the impact. The superscript $loc$ indicates whether the count is for \textit{in location} or \textit{out location} tweets.

To fairly assess changes in the online attention to climate change, we compare the aggregated tweet count to two different baselines. The first baseline, referred to as the ``random baseline'', is the count of the number of tweets on 100 randomly selected dates within the time interval from January 1st 2010 to December 31st 2021. This baseline provides a general understanding of the climate change debate worldwide over the last 12 years. Results are robust using an alternative \textit{extra hurricane} baseline: we select 15 non-hurricane dates to establish a baseline for tweet counts about climate change, by excluding a three-month period before and after each hurricane to avoid overlaps, see SI. 
All counts using both baselines are normalised by the average tweet count from the 30 days prior to the selected date. For a statistical comparison of the changes in tweet count, we use the Students' $T$ test \cite{pearson1894contributions}.

\section*{Results and Discussion}
\label{sec:results}

\subsection*{Analysis of hurricane news: content and terminology}
\label{sec:top_model_results}

In order to shed light on the topics discussed in relation to hurricanes, we study the spread of news media content on Twitter related to hurricanes. This clarifies the connection between natural disasters and the public perception of climate change. 

\subsubsection*{Hurricane-related news articles}
\label{sec:hurricanes_news}

\begin{figure}[ht!]
\centering
\includegraphics[scale=0.33]{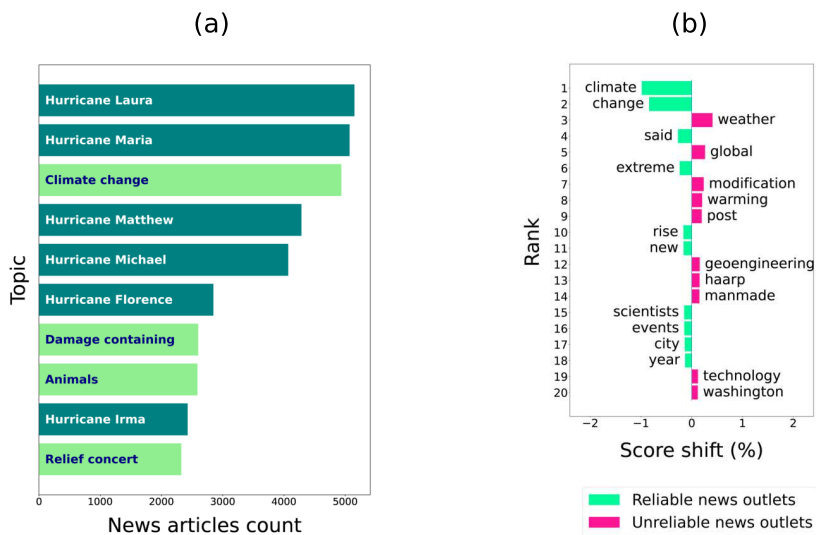}
\caption{\textbf{Leading topics in hurricane related news articles, and key news terminology by reliability of the news sources}. Climate change coverage is among the most covered topics in news articles about hurricanes. (a) The most prominent topics in the news dataset. Dark green bars correspond to hurricane specific topics, light green topics are not specific to an individual hurricane. (b) The terminology used in by reliable (green) and unreliable (purple) news articles in the climate change topic. Words are ranked in descending order by the relative frequency within the two sets. The score shift indicates whether the term is disproportionately used by reliable (left) or unreliable (right) news outlets.}
\label{fig:fig_all_bert_analyses}
\end{figure}

We now analyse the news articles referenced in the hurricane tweets to better understand the topics discussed during, and in the aftermath, of a hurricane impact.
To associate each article to a topic we train BERTopic on the hurricane news database, see Section \nameref{sec:top_mod}. Figure \ref{fig:fig_all_bert_analyses}(a) shows the top ten topics most covered by news articles in our dataset.

Figure \ref{fig:fig_all_bert_analyses}(a) shows that climate change is the leading topic which is not specific to an individual hurricane. Hurricane specific news typically provides  information on the regions impacted by the hurricane and the degree of damage caused. Of the set of climate related news identified, we find that around $15\%$ of news stories are from unreliable sources. Over time, the average NewsGuard score in our dataset is approximately stable, with little evidence that reliable or unreliable news sources are more prominent, see Fig SI2. 

We now analyse the relative importance of the terminology used by reliable and unreliable news sources by quantifying which words contribute to a pairwise difference between two texts. Figure \ref{fig:fig_all_bert_analyses}(b) shows that reliable news sources disproportionately refer to \say{climate change}, whereas unreliable news sources prefer terms such as \say{global warming} and \say{weather}. This finding aligns with previous research that has associated the term ``global warming'' with hoax frames and less scientifically accurate content \cite{jang2015polarized,andersen2013key}. Other terms used include \say{future}, \say{damages}, and \say{events}, which have a neutral connotation that reflects news with informative, factual content.

On the other hand, in unreliable news articles (magenta), we find references to terms that are often used by conspiracy theorists (such as \say{modification}, \say{geoengineering} or \say{haarp}) to suggest that governments, or other powerful entities, are manipulating the climate for their own benefit \cite{shepherd2009geoengineering}. These words sit in contrast to the terms used by reliable news sources such as \say{scientists}, \say{report} or \say{study}. The term \say{haarp} is of particular interest, referring to High-Frequency Active Auroral Research Program, the  US civil and military installation located in Alaska, discussed principally following hurricane Sandy \cite{krehm1999meltdown}.

Such conspiracy theories are important since they can influence public attitudes towards geoengineering \cite{debnath2023conspiracy}. These conspiracy theories have the potential to undermine trust in scientific experts and institutions, making it more difficult to build support for climate action. 

\subsection*{Hurricanes increase online discussion around climate change}

\label{sec:hurricanes_on_twitter}
We now assess whether hurricanes impacting a local region affect the public discussion of climate change. We do this using geolocated climate change tweets which, we note, do not necessarily refer to specific hurricanes. By using geolocalized tweets, our analysis distinguishes between the effect a hurricane has in the region impacted, and outside that region. 

We compare the distribution of geolocated tweets within the affected areas (the \textit{in location}) and outside (the \textit{out of location}) by normalising the tweet counts, see Section \nameref{sec:stat_an}. The change in tweet count is compared to two random baselines, see Section \nameref{sec:stat_an}. Statistical analysis comparing the in-location and out-of-location change in tweet count to the baselines is provided in Table SI1. 

\begin{figure}[ht!]
\centering
\includegraphics[scale=0.33]{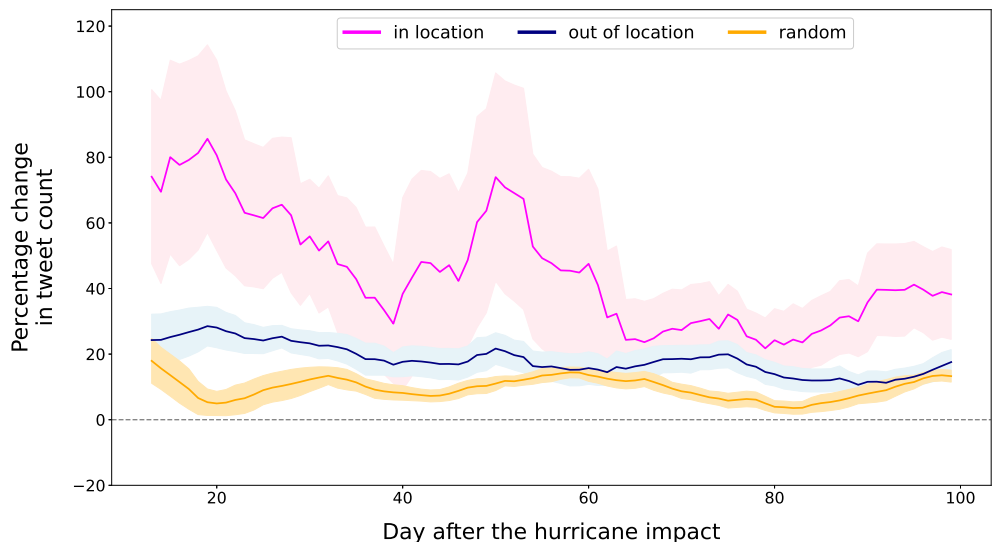}
\caption{\textbf{Hurricane impact on Twitter attention in the affected areas with respect to not affected areas.} We show the percentage change in the number of tweets after the hurricane, with respect to the average number of tweets in the 30 days before the hurricane. We compare the \textit{in location} (pink line) and the \textit{out of location} curve (blue line) with respect to the \textit{random} baseline (orange line). The shadow areas around the curves are the standard deviations of the mean among all the hurricanes.} 
\label{fig:fig_comparison_among_distributions_random_only_rolling_all}
\end{figure}

Figure \ref{fig:fig_comparison_among_distributions_random_only_rolling_all} shows the percentage change in the number of tweets after the hurricane impacts, with
respect to the average of the number of tweets in the 30 days before, for the \textit{in location}, \textit{out of location}, and random baseline. For each curve, the shaded area corresponds to the standard deviation of the percentage change in the aggregated average tweet count across all hurricanes.

The biggest positive increase in tweet count is for the \textit{in location} curve, an increase that peaks at around 80\% in the first three weeks following impact, before decaying to 40\% at the end of the three month follow-up period. We note that if we only analyse the largest hurricanes (measured in terms of USD damage) then \textit{in location} tweets related to climate change can increase by up to 200\%, see SI Fig SI4 and Fig SI5. 

The \textit{out of location} curve also increases following a hurricane's impact, with a stable increase in the three months following impact of around 20\%. In general, the percentage change in the number of tweets about climate change right after a hurricane is significant; the \textit{random} baseline in Fig \ref{fig:fig_comparison_among_distributions_random_only_rolling_all}, fluctuates but remains below 20\% throughout. We stress that the \textit{random} baseline may include dates that coincide with both the impact of hurricanes and other events related to climate change. As a result, any variations in the baseline data may be attributed to these factors. However, our analysis reveals that these variations are small.

We note that for the number of \textit{in location} tweets there is a reduction after a period of around 2 months. Indeed, looking at the percentage increase after 8 weeks, the values for the four categories (\textit{in location}, \textit{out of location}, \textit{random} and \textit{extra hurricanes}) are comparable. The comparisons of the above distributions using the Students' T-test are shown in Table SI1. Based on the statistical test conducted, the results show that the \textit{in location} distribution is significantly larger than the \textit{out of location}, \textit{random}, and \textit{extra hurricane} baselines. The \textit{out of location} curve is also found to be significantly larger than both baselines.

\section*{Conclusions}
\label{sec:concl}

In this paper, we explored the impact of hurricanes on the public attention towards climate change over the past 12 years. With respect to the previous literature on hurricane impacts and social media \cite{roxburgh2019characterising}, we studied a wider range of hurricanes, placing a particular emphasis on the spatial and temporal effects of a hurricane's impact, and considering the reliability of news media sources when analysing content. 

Our analysis shows that hurricanes trigger a surge in the online discussion on climate change, as indicated by the increased use of climate change related terms in tweets and news articles following a hurricane. In regions affected by a hurricane, the number of climate change related tweets increases by 80\% after impact, and up to 200\% for the largest hurricanes.
Note, however, that such an increase is limited both temporally and geographically, with a rapid decay in the public attention towards climate change. Our findings imply that the heightened public concern and focus towards climate change might have a transient nature, highlighting the necessity for ongoing endeavors to ensure continuous public engagement with the issue, extending beyond the immediate aftermath of a natural disaster.

With regards to the news coverage about hurricanes, the choice of terminology can signal the reliability of a news source and how it chooses to frame discussions around climate change \cite{jang2015polarized,andersen2013key,falkenberg2022growing}.

In accordance with the findings of our study, it has been observed that trustworthy news outlets are more inclined to use the phrase ``climate change'' in their publications, whereas less credible sources have a tendency to favour the terms ``global warming'' and ``weather'' \cite{jang2015polarized}, \cite{andersen2013key}. Furthermore, references to "HAARP" –- a conspiracy theory asserting the US government's manipulation of weather through a radio transmitter –- in untrustworthy media sources underscores the capacity of Twitter to disseminate climate-related conspiracy theories and misinformation \cite{debnath2023conspiracy}.

Media outlets with a climate sceptic agenda often prefer language which emphasises uncertainties in the science; such language comes under the broader set of themes often referred to as the ``discourses of delay'' \cite{lamb2020discourses}. Identifying such claims and studying their spread is becoming increasingly important given that recent evidence has shown that particular discourses related to political hypocrisy and inaction can offer a gateway into climate sceptic communities on social media for regular users \cite{falkenberg2022growing}. 

There are limitations to our study which present opportunities for future work. First, our social media analysis is limited to Twitter. Previous work suggests that Twitter dominates other social media platforms in the online discussion around climate change \cite{falkenberg2022growing}, however, future work should consider how climate change is communicated on other platforms. Second, we have restricted our analysis to hurricanes when a discussion of other extreme weather events (e.g., droughts, floods, heatwaves) would be equally warranted. However, accurately retrieving data for such events which, unlike hurricanes, are not uniquely named is difficult. Third, our analysis only considers English language tweets referring to tropical storms in North America. Future work should consider storms in other parts of the world and should analyse non-English language content. Finally, our datasets are keyword-based which miss part of the relevant discussion around climate change and hurricanes. This is a common limitation in most Twitter-based communication studies, but extending analysis to a broader set of terms could be beneficial for furthering our analysis.

The results of our study hold significance for policymakers who aim to efficiently tackle climate change and curb the spread of climate misinformation. Specifically, the transient spike in climate change awareness that occurs in the aftermath of a hurricane indicates that efforts to counteract climate misinformation should be implemented proactively. Interventions activated only in the weeks following a hurricane may not garner the same level of attention as those executed immediately after a hurricane's impact.

In summary, this research offers valuable insights as to how hurricanes influence the public's focus on climate change and emphasizes the need for continuous endeavors to preserve engagement with this critical subject beyond the immediate consequences of a natural disaster.

\section*{Acknowledgements}
M.T., M.F., A.G., F.Z., W.Q. and A.B. acknowledge support from the IRIS Infodemic Coalition (UK government, grant no. SCH-00001-3391). F.Z. acknowledges financial support from the European Union’s Rights, Equality and Citizenship project EUMEPLAT grant no. 101004488.

\bibliography{biblio}

\end{document}